\documentclass[12pt,preprint]{aastex}

\begin{document}


\title{The Average Size and Temperature Profile of Quasar Accretion Disks}


\author{J. Jim\'enez-Vicente}
\affil{Departamento de F\'{\i}sica Te\'orica y del Cosmos. Universidad de Granada, Campus de Fuentenueva, 18071, Granada, Spain \\
Instituto Carlos I de F\'{\i}sica Te\'orica y Computacional. Universidad de Granada, 18071, Granada, Spain}
\author{E. Mediavilla}
\affil{Instituto de Astrof\'{\i}sica de Canarias, V\'{\i}a L\'actea S/N, La Laguna 38200, Tenerife, Spain \\
Departamento de Astrof\'{\i}sica, Universidad de la Laguna, La Laguna 38200, Tenerife, Spain}
\author{C. S. Kochanek}
\affil{Department of Astronomy, The Ohio State University, 140 West 18th Avenue, Columbus, OH 43210, USA\\
Center for Cosmology and Astroparticle Physics, The Ohio State University, 191 West Woodruff Avenue, Columbus, OH 43210, USA}
\author{J. A. Mu\~noz}
\affil{Departamento de Astronom\'{\i}a y Astrof\'{\i}sica, Universidad de Valencia, 46100 Burjassot, Valencia, Spain}
\author{V. Motta}
\affil{Departamento de F\'{\i}sica y Astronom\'{\i}a, Universidad de Valpara\'{\i}so, Avda. Gran Breta\~na 1111, Playa Ancha, 
Valpara\'{\i}so 2360102, Chile}
\author{E. Falco}
\affil{Harvard-Smithsonian Center for Astrophysics, Cambridge, MA 02138, USA} 
\author{A. M. Mosquera}
\affil{Department of Astronomy, The Ohio State University, 140 West 18th Avenue, Columbus, OH 43210, USA
}




\begin{abstract}

We use multi-wavelength microlensing measurements of a sample of 10 image pairs from 8 lensed
quasars to study the structure of their accretion disks. By using
spectroscopy or narrow band photometry we have been
able to remove contamination from the weakly microlensed
broad emission lines, extinction and any uncertainties in the large-scale macro
magnification of the lens model. We determine a maximum likelihood estimate for the exponent of
the size versus wavelength scaling ($r_s\propto \lambda^p$ corresponding
to a disk temperature profile of $T\propto r^{-1/p}$) of $p=0.75^{+0.2}_{-0.2}$, and a Bayesian estimate of $p=0.8\pm0.2$, which are 
significantly smaller than the prediction of thin disk theory ($p=4/3$).  
We have
also obtained a maximum likelihood estimate for the average quasar accretion disk size
of $r_s=4.5^{+1.5}_{-1.2} $ lt-day at a rest frame wavelength of $\lambda = 1026~{\mathrm \AA}$ 
for microlenses with a
mean mass of $M=1 M_\sun$, in agreement with previous results, and
larger than expected from thin disk theory.

\end{abstract}





\section{Introduction}

The standard model to describe the generation of energy in the nucleus of
quasars and AGN is the thin accretion disk (Shakura \& Sunyaev 1973; Novikov \&
Thorne 1973). This model not only explains the luminosities of 
quasars, but
also predicts observables such as the disk size and its dependence on
wavelength. 
In particular, the
scaling of the disk size with wavelength, or equivalently, 
the temperature profile of
the disk, is a fundamental test for any disk theory.
Well away from the inner edge, a temperature profile $T\propto r^{-1/p}$ translates
into a size profile $r_\lambda \propto \lambda^p$, where the characteristic size
can be thought of as the radius where the rest wavelength matches the disk
temperature. The simple Shakura \& Sunyaev (1973) disk has $T\propto r^{-3/4}$
and hence $p=4/3$.

One of the most effective means of measuring accretion disk sizes
is gravitational microlensing of lensed quasars (Wambsganss 2006).
There are now many quasar size measurements based on microlensing either of
individual objects (Kochanek 2004; Anguita et al. 2008; Eigenbrod et al. 2008;
 Poindexter et al. 2008; Morgan et
al. 2008; Bate et al. 2008; Mosquera et al. 2009; Agol et al. 2009; Floyd et al. 2009; Poindexter \& Kochanek 2010;  Dai et al. 2010;
Mosquera et al. 2011; Mediavilla
et al. 2011a; Mu\~noz et al. 2011;  Blackburne et al. 2011b, Hainline
et al. 2012; Motta et al. 2012; Mosquera et al. 2013; Blackburne et al. 2013) or of samples of lenses (Pooley et al. 2007; Morgan et al. 2010; Blackburne
et al. 2011a, Jim\'enez-Vicente et al. 2012). A generic outcome of these
studies is that the disks are larger than predicted. 
One possible solution is to modify the disk temperature profiles,
and this can be tested using microlensing by measuring the
dependence of the disk size on wavelength.

There are fewer estimates of the wavelength dependence of the size (see Blackburne et al. 2011a and Morgan et al. 2010 and references therein).
Most of the measurements (Anguita et
al. 2008, Poindexter et al. 2008, Bate et al. 2008, Eigenbrod et al. 2008, Floyd
et al. 2009, Blackburne et al. 2011a, Blackburne et al. 2013) are based on broad
band photometry that can be affected by line contamination. 
These studies generally find that the shorter wavelength emission regions are
more compact, but the estimates of the scaling exponent $p$ have both
large individual uncertainties and a broad scatter between studies.
Only Blackburne et al. (2011a) have found little or no dependence of size on
wavelength. 

Single epoch broad band photometry is affected by several potential systematic
problems: i) broad lines, which are known to be less affected by microlensing both observationally
(e.g. Guerras et al. 2013a) and theoretically (e.g. Abajas et al. 2002), ii) differential extinction
of the images by the lens galaxy (e.g. Falco et al. 1999, Mu\~noz et al. 2004), and iii) uncertainties in the macro model lens magnification.
The effects of extinction and modeling uncertainties can be eliminated by analysing the 
time variability produced by microlensing, but the effects of line contamination
must still be modeled (e.g. see the Dai et al. 2010 analysis of RXJ1131$-$1231). We can remove the effect of broad line contamination by using 
spectroscopy (Mediavilla et al. 2011a, Motta et al. 2012) or narrow band
photometry (Mosquera et al. 2009, Mosquera et al. 2011) to measure
the continuum between the lines. If the continuum and line flux ratios can be separately measured, then the line flux
ratios can also be used to eliminate the effects of extinction and model uncertainties on the continuum flux
ratios to the extent that the lines are little affected by microlensing.

Even with spectra, there are residual issues coming from line contributions to the apparent continuum.
The most important of these are the emission from FeII in the wavelength range from 1800 to 3500 \AA\ 
 and the Balmer continuum in the range from 2700 to 3800 \AA\ (Wills, Netzer \& Wills 1985).
The contribution of these pseudo-continua to the measured continuum flux varies
from object to object, and its effect on the estimate of the continuum microlensing magnifications is 
difficult to assess.
If the UV emitting Fe II comes from a region similar in size to the continuum 
source as suggested by Guerras et al. (2013b), it would
be similarly affected by microlensing. On the other hand, the Balmer continuum 
likely forms on larger scales and would be
less affected by microlensing (see Maoz et al. 1993). 
Like including a broad line, it would dilute the microlensing of the continuum. Since the 
Balmer continuum peaks near the Balmer limit at 3646 \AA,
it primarily affects the redder portion of the wavelength ranges we consider here
and so could slightly bias our results towards steeper slopes (higher $p$).
The scale of the effect is difficult to estimate, but it should be modest. 

Previous results based
on spectroscopic or narrow band studies (SBS0909+532 in Mediavilla et al. 2011a, 
SDSSJ1004+4112 in Motta et al. 2012, HE1104$-$1805 in Motta et al. 2012 and
in Mu\~noz et al. 2011, HE0435$-$1223 in
Mosquera et al. 2011) have found a
slightly shallower wavelength dependence than the 
predictions of the standard thin disk with $p\approx 1$ rather than $p=4/3$.
The main objective of the present work is to extend the studies
of the exponent of the size-wavelength scaling, $p$ ($r_s\propto \lambda ^p$),
using spectroscopic and narrow band data that are less affected by the presence of emission
lines or extinction (\S2). In \S3 we will calculate an estimate for the average
value of the exponent $p$ based on this less line contaminated data. 
The main conclusions are
presented in \S4.

\section{Data Analysis}

In the present study we use estimates of
microlensing magnifications measured either from optical/infrared spectra or 
narrowband photometry available in the literature. 
We collected data for the 8 lens systems in Table 
\ref{tab1}.
Spectroscopic data are available for HE0512$-$3329 (Wucknitz et al., 2003),
SBS0909+532 (Mediavilla et al. 2011), QSO0957+561, SDSSJ1004+4112 and
HE1104$-$1805 (Motta et al. 2012). In the latter case, 
there were two epochs of data showing significant changes in the
microlensing that were examined in Motta et al. (2012).
We also selected objects with narrowband photometry for which (most of) 
the observed filters
do not contain contamination from strong emission lines and
where there is little extinction or where we can make extinction correction. 
This adds two more objects: HE0435$-$1223 
(Mosquera et al. 2011) and Q2237+0305 
(Mu\~noz et al. 2014).

For spectroscopic data, the microlensing magnification between two images, 
1 and 2, at a given wavelength $\lambda$
is calculated as $\Delta m = (m_2 - m_1)_{micro} = (m_2 - m_1)_{cont} - (m_2 -
m_1)_{line}$, where $(m_2 - m_1)_{line}$ is the flux ratio in an 
emission line
and $(m_2 - m_1)_{cont}$ is the flux ratio of the continuum adjacent to that
emission line. With this method, the microlensing magnification is 
isolated from extinction and from the mean lensing magnification, 
as these affect the line and
the adjacent continuum equally (Mediavilla et al. 2009). 
With narrow band photometry
we can also isolate the microlensing in the continuum from the contamination 
of strong emission lines. 
In this case, however, additional information from IR or spectroscopic
observations may be needed to safely remove extinction (see Mosquera et al. 2011,
Mu\~noz et al. 2011) and to provide an unmicrolensed baseline. 

For each object (except Q2237+0305 for which Mu\~noz et al. 2014 carried out a 
detailed
analysis of multi-epoch data in five different bands), 
we used the measured microlensing magnifications at three 
different wavelengths, trying to keep the wavelength baseline as large as
possible in order to maximize our sensitivity to chromatic microlensing. 
Therefore, we used the
bluest and reddest ends of the observed ranges, with a third point at an 
intermediate wavelength, using data from the literature as summarized in
Table \ref{tab1}. In some cases the microlensing
magnification at the three wavelengths is
directly obtained from the original references. In other cases, the values
where calculated from the published line and continuum magnitude differences. 
For HE0435$-$1223, for which we have narrowband photometry, 
the continuum magnifications were taken from the observations and
the H band magnifications were used as an
unmicrolensed baseline (Mosquera et al. 2011).
Similarly, mid-infrared flux ratios from Minezaki et al. (2009) were used
as an unmicrolensed baseline for Q2237+0305. There may be
some residual differential extinction in Q2237+0305, but
it is very small compared to the measured microlensing chromaticity
and can be ignored (see Mu\~noz et al. 2014). 
The values of the differential microlensing magnification at each wavelength 
and their estimated errors are shown in 
Table \ref{tab1} along with the definition of the
unmicrolensed baseline and the source of the data.

\begin{deluxetable}{llrrrc}
\tabletypesize{\footnotesize}
\tablewidth{0pt}
\rotate
\tablecolumns{6}
\tablecaption{Chromatic microlensing magnifications\label{tab1}}
\tablehead{
\colhead{Object} & \colhead{Pair} & \colhead{$\Delta m @ \lambda_1 $} &
\colhead{$\Delta m @ \lambda_2 $} & \colhead{$\Delta m @ \lambda_3 $} &
\colhead{\shortstack{Unmicrolensed \\ Baseline}}
}
\startdata 
HE0435$-$1223\tablenotemark{a} & B$-$A &   0.38$\pm$0.05 @ 1305 \AA\   & 0.32$\pm$0.05 @ 1737 \AA\   & 0.15$\pm$0.05 @ 2815 \AA\   & NIR \\
HE0512$-$3329\tablenotemark{b} & B$-$A & $0.73\pm$0.12 @ 1026 \AA\   & $0.33\pm$0.11 @ 1474 \AA\   & $0.34\pm$0.06 @ 1909 \AA\   &  Lines \\
SBS0909+532\tablenotemark{c} & B$-$A & $-0.67\pm$0.05 @ 1459 \AA\   & $-0.30\pm$0.10 @ 4281 \AA\   & $-0.24\pm$0.07 @ 6559 \AA\   &  Lines \\ 
QSO0957+561\tablenotemark{d} & B$-$A & $-0.47\pm$0.09 @ 1216 \AA\   & $-0.45\pm$0.09 @ 1909 \AA\   & $-0.41\pm$0.09 @ 2796 \AA\   & Lines \\ 
SDSSJ1004+4112\tablenotemark{d}& B$-$A & $-0.60\pm$0.07 @ 1353 \AA\   & $-0.40\pm$0.07 @ 2318 \AA\   & $-0.08\pm$0.07 @ 4572 \AA\   & Lines \\
SDSSJ1029+2623\tablenotemark{d}& B$-$A & $0.30\pm$0.10 @ 1216 \AA\   & $0.42\pm$0.10 @ 1549 \AA\   & $0.47\pm$0.10 @ 1909 \AA\   & Lines \\
HE1104$-$1805$\tablenotemark{e}$ & B$-$A & 0.65$\pm$0.07 @ 1672 \AA\    & $0.46\pm$0.07 @ 2452 \AA\   & 0.32$\pm$0.07 @ 4669 \AA\   & Lines \\
HE1104$-$1805$\tablenotemark{f}$ & B$-$A & $-0.03\pm$0.07 @ 1306 \AA\    & 0.05$\pm$0.07 @ 1728 \AA\   & 0.18$\pm$0.07 @ 2796 \AA\   & Lines \\
Q2237+0305\tablenotemark{g} & A$-$D, B$-$D, C$-$D & & & & MIR \\
\enddata
\tablecomments{Wavelengths are indicated in the rest frame.}
\tablenotetext{a}{From Mosquera et al. (2011)}
\tablenotetext{b}{From Wucknitz et al. (2003)}
\tablenotetext{c}{From Mediavilla et al. (2011)}
\tablenotetext{d}{From Motta et al. (2012)}
\tablenotetext{e}{Pre-2003 observations. From Motta et al. (2012)}
\tablenotetext{f}{Post-2006 observations. From Motta et al. (2012)}
\tablenotetext{g}{From Mu\~noz et al. (2014). For this object, the analysis is based on six epochs of data five at five wavelengths (narrow band filters) 
for the three independent image pairs. This data are reported in Mu\~noz et al. (2014).}
\end{deluxetable}

We used a Monte Carlo statistical analysis to estimate the average size and its
wavelength dependence for this sample of objects. The procedure is very similar
to the one described in Jim\'enez-Vicente et al. (2012), but in 
this case we apply it to the estimates of the
microlensing magnification at three different wavelengths instead
of two. The structure of the accretion disk is described by a Gaussian
profile $I(R)\propto \exp(-R^2/2r_s^2)$, where the characteristic size $r_s$
is a wavelength-dependent parameter (related to the half-light
radius by $R_{1/2}=1.18r_s$). The wavelength dependence of the source
size is parametrized by a power law such
that $r_s(\lambda) \propto \lambda^p$, and as noted earlier, the standard 
Shakura \& Sunyaev (1973) accretion disk model has $p=4/3$.

Magnification maps for each image
are generated using the Inverse Polygon Mapping (IPM) algorithm (Mediavilla et al. 2006, 2011b). 
The surface densities near the lensed images are generally dark matter dominated, so 
we put 5\% of the surface mass density in the form of stars following the
estimates of Mediavilla et al. (2009). 
The one exception was Q2237+0305, where 
100\% of the mass is in 
form of stars because the lensing is dominated by the bulge
of a low redshift spiral galaxy. We used a fixed microlens mass of $M=M_\sun$.
All linear scales 
can be rescaled with the square root of
the microlens mass ($r_s(\lambda)\propto(M/M_\sun)^{1/2}$).
The values of the convergence $\kappa$ and shear $\gamma$ associated with the macro
lens models were 
taken from Mediavilla et al. (2009). For SBS0909+532,
the improved values in Mediavilla et al. (2011b) were used.
The maps are 2000$\times$2000 pixels in size, with a pixel size of
0.5 lt-days. 
For the objects in our sample 
this size corresponds to between 43 and 80 Einstein radii 
(with the exception of Q2237+0305,
for which the map size is 14 Einstein radii). 
We explored a grid in the parameter space of
 the logarithmic slope $p$ and the disk size 
at a reference wavelength $r_s=r_s(\lambda_0)$. The reference
wavelength is taken to be the bluest wavelength
for which we have a measured microlensing magnification, which
is 1026\AA\ (rest frame). 
The grid runs over $p=0.25\times i$ for $i=0\cdots 10$.
We use a natural logarithmic grid in $r_s$ such that
$\ln r_s=0.3\times j$ for $j=0\cdots 11$. 

For every pair of values $(p,r_s)$, the magnification 
maps are convolved with Gaussians of the relevant sizes at the
three wavelengths $r_s(\lambda_k)=r_s(\lambda_k/\lambda_0)^p$ with
$k=1,2,3$ from which we can compare the microlensing
magnification statistics with the observed values at those wavelengths.
The likelihood of observing the three microlensing magnifications
$\Delta m^{obs}_k$ for lens $l$ at these three wavelengths $\lambda_k$ ($k=1,2,3$) given
the parameters $p$ and $\ln(r_s)$ is  calculated as
\begin{equation}
\label{eq1}
P_l(\Delta m^{obs}_k | p_i,\ln(r_{sj})) \propto
\int \Delta m_1 \int \Delta m_2 \int \Delta m_3 N_{ij} e^{-\frac{1}{2} \chi^2}
\end{equation}
where $N_{ij}$ is the number of trials with $\Delta m_1$, $\Delta m_2$ and $\Delta m_3$ for the case with parameters $p_i$ and, $\ln(r_{sj})$, and
\begin{equation}
\chi^2=\sum_{k=1}^{3} \frac{(\Delta m^{obs}_k-\Delta m_k)^2}{\sigma_k^2}.
\end{equation}
The estimated errors  $\sigma_k$ at the different wavelengths are given in
Table \ref{tab1}.
The likelihood given by Equation \ref{eq1} is calculated
for each of the 8 lenses shown in Table \ref{tab1}\footnote{The case
of Q2237+0305 is calculated as described in Mu\~noz 
et al. (2014)}. 
For each value of the pair $(p,\ln(r_s))$, the 
likelihood is calculated using $10^8$ trials 
by sampling each of the two magnification maps at 
$10^4$ points distributed on a regular grid.
We construct a joint likelihood function for the parameters $p$ and $\ln(r_s)$
by multiplying the individual likelihood functions for the 8 lenses
\begin{equation}
P(\Delta m^{obs}_{k,l} | p_i,\ln(r_{sj}))\propto\prod_{l=1}^{8} P_l(\Delta m^{obs}_k | p_i,\ln(r_{sj}))
\end{equation}  

These likelihoods are shown in Figures 1 (for the individual objects) and 
2 and 3 (for the whole sample). They can be directly used to estimate values
for the parameters $p$ and $\ln(r_s)$. In this context we have used the 
Maximum Likelihood Estimator.
We can also take a Bayesian approach and convert these 
likelihoods into posterior probabilities by multiplying them by suitable
priors. We have used uniform priors for $\ln(r_s)$ (equivalent to a 
logarithmic prior for $r_s$) and
the exponent $p$. With this choice for the priors,
Figures 1, 2 and 3 also represent the Bayesian
posterior probabilities (normalization constants aside). 
In this approach, the mean value of the posterior
probabilities and its dispersion are used as estimators of the parameters.

\section{Results and Discussion}

\begin{figure}
\epsscale{0.8}
\plotone{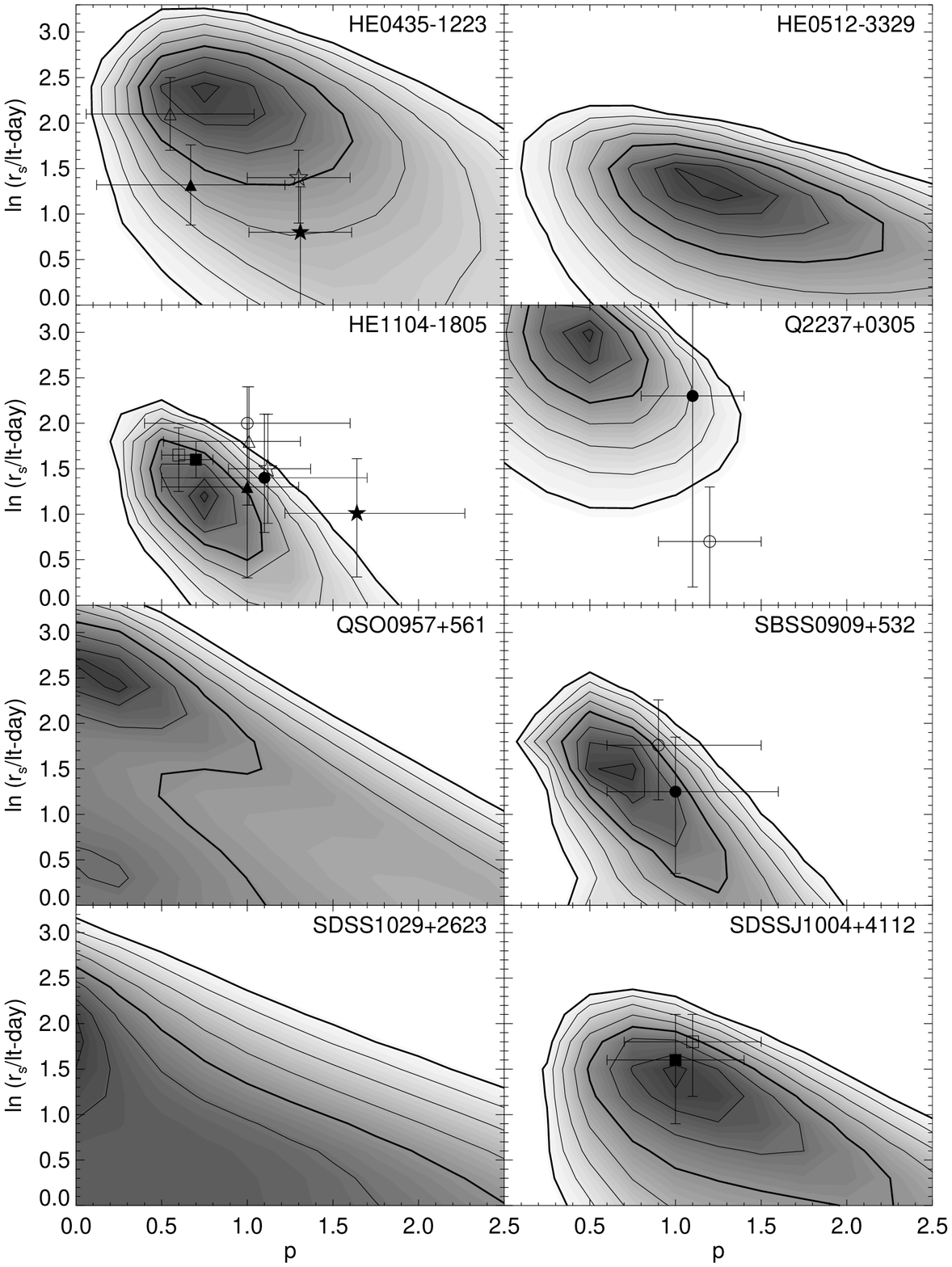}
\caption{Likelihood distributions for the size of the accretion disk, ${\ln}(r_s/\mathrm{lt-day})$ at 1026 \AA, and
the logarithmic slope of the size $p$ ($r_s\propto \lambda^p$) for each 
individual lens systems. 
The contour levels are drawn at 
likelihood intervals of
0.25$\sigma$  for one parameter 
from the maximum.
The contours at 1$\sigma$ and 
2$\sigma$ are heavier. Published results rescaled to our baselines 
are overplotted as open (filled) symbols for estimates using linear (logarithmic) priors.
Triangles are from Blackburne et al. (2011a) for HE0435$-$1223 and from Blackburne et al. (2013) for HE1104$-$1805.
Stars are from Mosquera et al. (2011) for HE0435$-$1223 and from Poindexter et al. (2008) for HE1104$-$1805.
Circles are from Mediavilla et al. (2011) for SBS0909+532, from Mu\~noz et al. (2011) for HE1104$-$1805 and from
Eigenbrod et al. (2008) for Q2237+0305. In Q2237+0305 the open (filled) symbol is the estimate with (without) velocity prior.
Squares are from Motta et al. (2012).\label{fig1}}
\end{figure}
\begin{deluxetable}{lrrrrlll}
\tabletypesize{\footnotesize}
\tablewidth{0pt}
\tablecaption{Disk Parameter Estimates\label{tab2}}
\tablehead{ & \multicolumn{2}{c}{Max. Likelihood} & \multicolumn{2}{c}{Bayesian} & & &\\
\colhead{Object} & \colhead{$r_s$ (lt-day)} & \colhead{$p$} & \colhead{$r_s$ (lt-day)} & \colhead{$p$} & $\frac{M_{BH}}{10^9 M_\sun}$ & Line & Source }
\startdata 
HE0435$-$1223 & $11.0^{+5.4}_{-6.5}$ &  $ 0.75^{+0.8}_{-0.4}$ & $4.8^{+6.2}_{-2.7}$ &  $ 1.3\pm0.6$ & 0.50 & CIV & Peng et al. (2006) \\
HE0512$-$3329 & $3.3^{+2.7}_{-1.5}$ & $1.25^{+0.6}_{-0.7}$ &  $2.6^{+1.9}_{-1.1}$ & $1.4\pm0.6$ &- & - & -\\
SBS0909+532 & $4.5^{+3.6}_{-3.3}$ & $0.75^{+0.5}_{-0.4}$ &  $2.7^{+2.6}_{-1.3}$ & $0.9\pm0.4$ & 1.95 & ${\mathrm H}\beta$ & Assef et al. (2011) \\
QSO0957+561 & $11.0^{+11}_{-11}$ & $0.25^{+0.9}_{-0.2}$ & $3.5^{+5.1}_{-2.1}$ & $0.9\pm0.7$ & 0.72 & ${\mathrm H}\beta$ & Assef et al. (2011) \\
SDSSJ1004+4112& $4.5^{+2.9}_{-4.5}$ & $1.00^{+1.3}_{-0.5}$ & $2.5^{+2.1}_{-1.1}$ & $1.3\pm0.6$ & 2.02 & CIV & Peng et al. (2006) \\
SDSSJ1029+2623& $6.0^{+7.4}_{-6.0}$ & $0.00^{+2.5}$ & $2.6^{+3.0}_{-1.4}$ & $0.9\pm 0.7$ & - & - & - \\
HE1104$-$1805 & $3.3^{+2.7}_{-1.7}$ & $0.75^{+0.3}_{-0.3}$ &  $2.6^{+2.0}_{-1.1}$ & $0.9\pm 0.4$ & 0.59 & ${\mathrm H}\beta$ & Assef et al. (2011) \\
Q2237+0305 & $20.0^{+7.0}_{-10}$ & $0.50^{+0.3}_{-0.4}$  & $10.0^{+12}_{-5.5}$ & $0.6\pm 0.4$  & 1.20 & ${\mathrm H}\beta$ & Assef et al. (2011)\\
\enddata
\tablecomments{Maximum Likelihood (cols 2-3) and Bayesian (cols 4-5) estimates of sizes and scaling exponents $p$ ($r_s\propto \lambda^p$)
for individual objects. Confidence intervals are at 1$\sigma$ level
for one degree of freedom for the Maximum Likelihood estimates
and marginalized for the second variable for the Bayesian estimates. 
Mass of the central black hole, 
line used in the estimate and source (cols 6-8).
Sizes were estimated at 1026 \AA\ (rest frame) for microlenses of mass $M=1M_\sun$.}
\end{deluxetable}
\begin{figure}
\epsscale{1.0}
\plotone{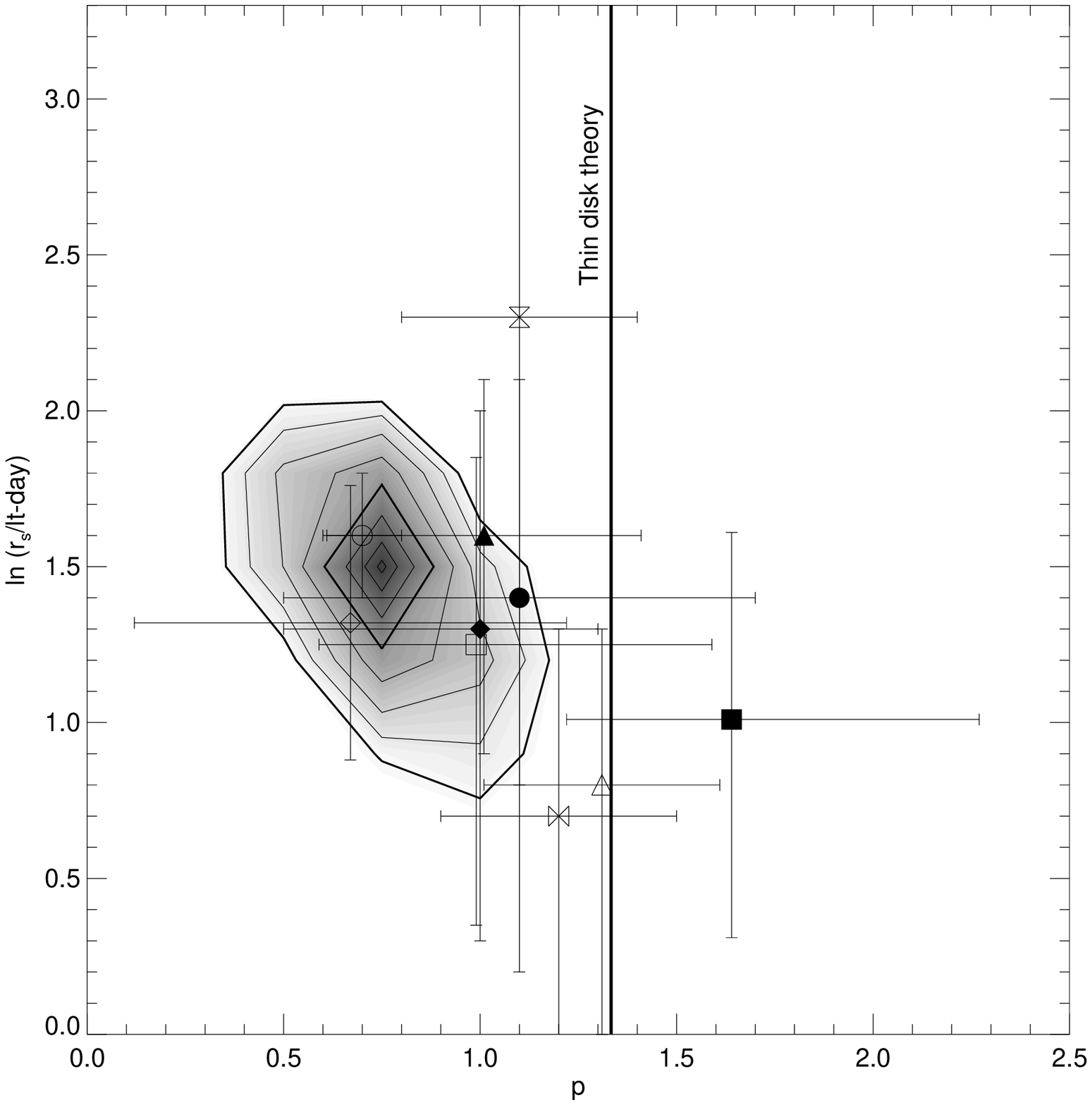}
\caption{Joint likelihood function for the size of the accretion disk 
at 1026 \AA, ${\ln}(r_s/\mathrm{lt-day})$, and
the logarithmic slope, $p$, of its scaling with wavelength ($r_s\propto\lambda^p$). The contour levels are drawn at intervals of
0.25$\sigma$ as in Fig. \ref{fig1}. Previous results are shown for 
Blackburne et al. (2011a, HE0435$-$1223, open diamond),
Mosquera et al. (2011, HE0435$-$1223, open triangle),
Motta et al. (2012, HE1104$-$1805, open circle), 
Blackburne et al. (2013, HE1104$-$1805, filled diamond), 
Poindexter et al. (2008, HE1104$-$1805, filled square), 
Mu\~{n}oz et al. (2011, HE1104$-$1805, filled circle),
Mediavilla et al. (2011, SBS0909+532, open square), 
Eigenbrod et al. (2008, Q2237+0305, bowtie/hourglass for estimate with/without velocity prior), 
Motta et al. (2012, SDSSJ1004+4112, filled triangle).
The vertical line indicates the prediction of the standard thin disk model, $p=4/3$.  \label{fig2}}
\end{figure}
The individual likelihood distributions for each of the 8 objects are 
shown in Figure \ref{fig1}.
For the case of HE1104$-$1805, where there is data at 
two different epochs, the distribution
is the product of the probability distributions for the two epochs.
Q2237+0305 combines the results for the three independent pairs 
at six different epochs (see Mu\~noz et al. 2014).
In Figure \ref{fig1} we also show the results from previous studies. In order to make this comparison, these sizes from the literature have, as necessary, been converted to a Gaussian $r_s$ at our nominal
wavelength of 1026 \AA\  using the reported values of $p$, a microlens
mass of $M=1M_\sun$, and assuming a fixed half-light radius, because 
estimates of the half-light radius are
largely model independent (Mortonson et al. 2005). The general agreement is very good, with our results being compatible with previous estimates
given the uncertainties. The maximum likelihood and Bayesian estimates for
 the size of the 
accretion disk $r_s$ at the reference wavelength
(1026 \AA) and for the chromaticity exponent $p$ are given in Table \ref{tab2}.
Figure \ref{fig2} shows the resulting joint likelihood. 
Previous estimates are also shown, although only for estimates
using logarithmic priors on size to avoid overcrowding. Again, the general agreement is very good.
The maximum likelihood estimates are
$p=0.75^{+0.2}_{-0.2}$ and $\ln(r_s/\mathrm{lt-day})=1.5^{+0.3}_{-0.3}$, where these are 1$\sigma$ confidence intervals for 1 parameter.
This corresponds to a  size  of $r_s(1026\ {\mathrm \AA})=4.5^{+1.5}_{-1.2}$ lt-days for a microlens mass of 1 $M_\sun$.
We repeated the calculations using only the objects with spectroscopic
data (i.e. excluding HE0435$-$1223 and Q2237+0305) to check whether
including narrowband photometry introduces any bias,
but the results are nearly identical. The most noticeable effect is that,
without these objects (mostly due to the effect of Q2237+0305), 
the distribution is slightly more extended towards
larger values of $p$ and smaller sizes. The upper limit on $p$ increases
by a small amount, resulting in an estimate of $p=0.75^{+0.3}_{-0.2}$.
In order to compare the present result for the size with 
recent estimates, we convert
it to a reference wavelength of 1736 \AA\ (using $p=0.75$) and
for a mass of the microlenses of 0.3 $M_\sun$.
The resulting (average) accretion disk size is $3.7^{+1.2}_{-1.0}$ lt-days, which is in good agreement with the
value of $4.0$ lt-days  estimated by Jim\'enez-Vicente et al. (2012)
from microlensing of (basically) this lens sample without chromatic information.

\begin{figure}
\epsscale{1.0}
\plotone{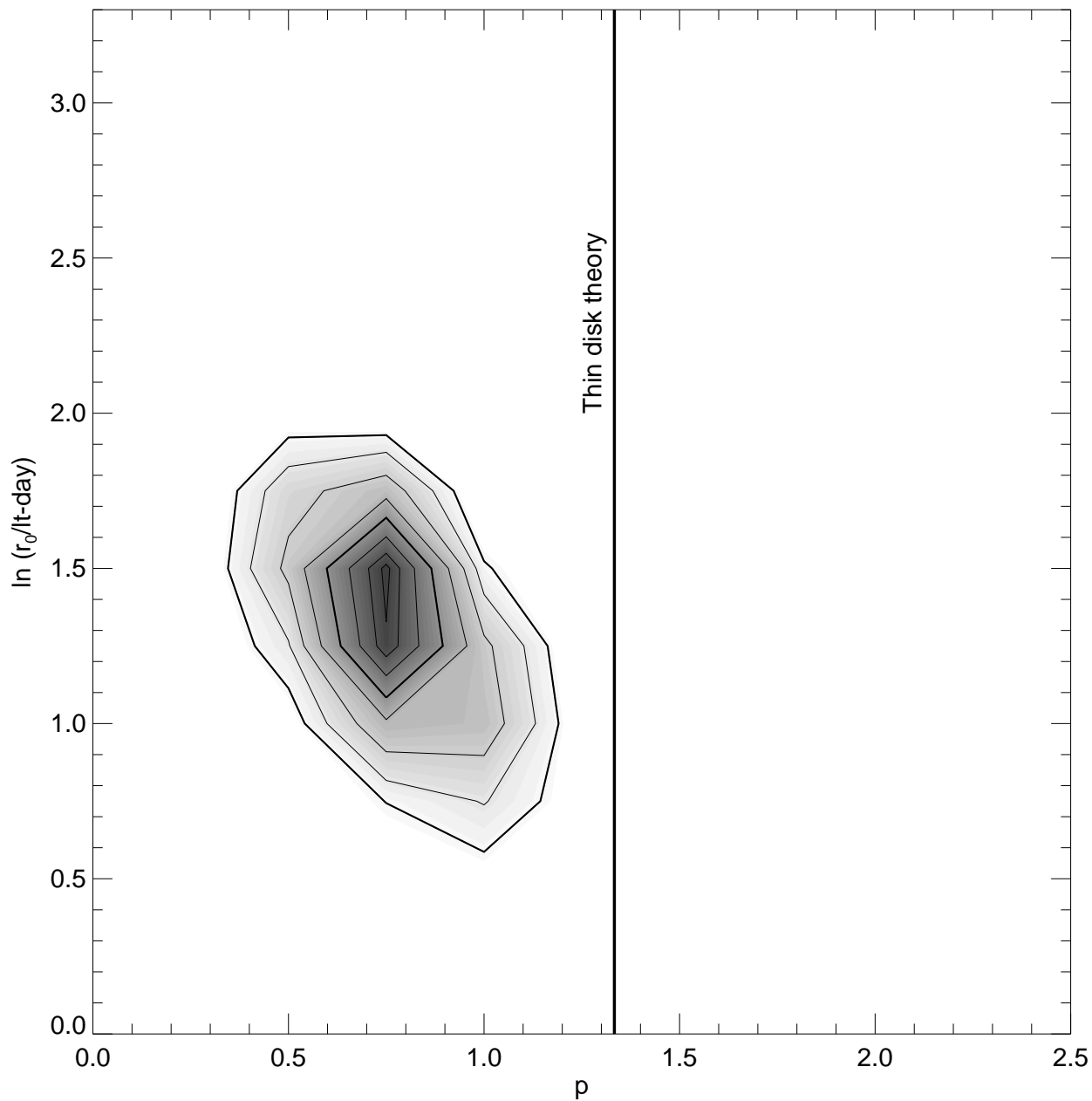}
\caption{Joint likelihood function for the size of the accretion disk at $M_{BH}=10^9 M_\sun$, ${\ln}(r_0/\mathrm{lt-days})$, and
the logarithmic slope, $p$, of its scaling with wavelength ($r_s\propto\lambda^p$). The contour levels are drawn at intervals of
0.25$\sigma$ as in Fig. \ref{fig1}. The vertical line indicates the prediction of the standard thin disk model, $p=4/3$. \label{fig3}}
\end{figure}

The result for the size is, however, an average value for
objects which, in principle, must have a spread in their
sizes due to differences in their black hole masses. 
In order to minimize this effect when combining
the likelihoods of different objects, we 
repeated the calculations scaling the reference size of the disk as
\begin{equation}
r_s(M_{BH})=r_0\left( \frac{M_{BH}}{10^9 M_\sun}\right)^{2/3}
\end{equation}
The 2/3 exponent is expected for Shakura \& Sunyaev disks radiating close to the Eddington limit, and
Morgan et al. (2010) used microlensing to estimate this scaling with mass and found
results consistent with the theoretical prediction. 
The black hole mass estimates are based on source luminosities and emission line widths.
We use estimates from the literature as reported in Table \ref{tab2}, using the $\mathrm{H}\beta$
estimates where available. Generally, these black hole mass 
estimates are viewed as logarithmically reliable with
uncertainties of factors of 2-3.
The resulting joint likelihood for the parameters $\ln(r_0)$ and $p$ is shown in Figure \ref{fig3}.
The maximum likelihood value is $\ln(r_0)=1.5^{+0.15}_{-0.45}$, corresponding to 
$r_0=4.5^{+0.7}_{-1.6}$ lt-days which is very similar to our previous result. 
This was
to be expected, as the average black hole mass for this sample is $1.2\times 10^9 M_\sun$, which is very close to the nominal mass used in the scaling. 
If we assume a typical value of $0.3 M_\sun$ for the mass of the
microlenses, our estimate for the size of the accretion disk
for a nominal black hole mass of $10^9 M_\sun$ 
is very similar to the estimate by Morgan et al. (2010).
The estimate of the logarithmic slope $p$ did not change,
remaining at $p=0.75^{+0.2}_{-0.2}$.

On scales of $r_0$, we would not expect a strong object 
dependence of the value of $p$ for objects sharing
the same emission mechanism.
In spite of this, there is no general agreement on the value of $p$ from
previous studies.
Most earlier determinations for individual objects seemed to point near the canonical value of $p=4/3$ (e.g. Eigenbrod et al., 2008, 
Poindexter et al. 2008, Floyd et al., 2009, Mosquera et al., 2011), with the notable exception of the sample 
analyzed by Blackburne et al. (2011a), which produced an unexpectedly low value of $p=0.17\pm0.15$ that is essentially compatible with a disk size that is independent
of wavelength. 
Nevertheless, several recent estimates based on different techniques 
(Mu\~noz et al., 2011, Mediavilla et al., 2011, Motta et al., 2012, Blackburne et al. 2013) have been finding
values close to $p=1$,
although with uncertainties large enough to be consistent with the canonical 
value of $p=4/3$. 
\begin{figure}
\epsscale{1.0}
\plotone{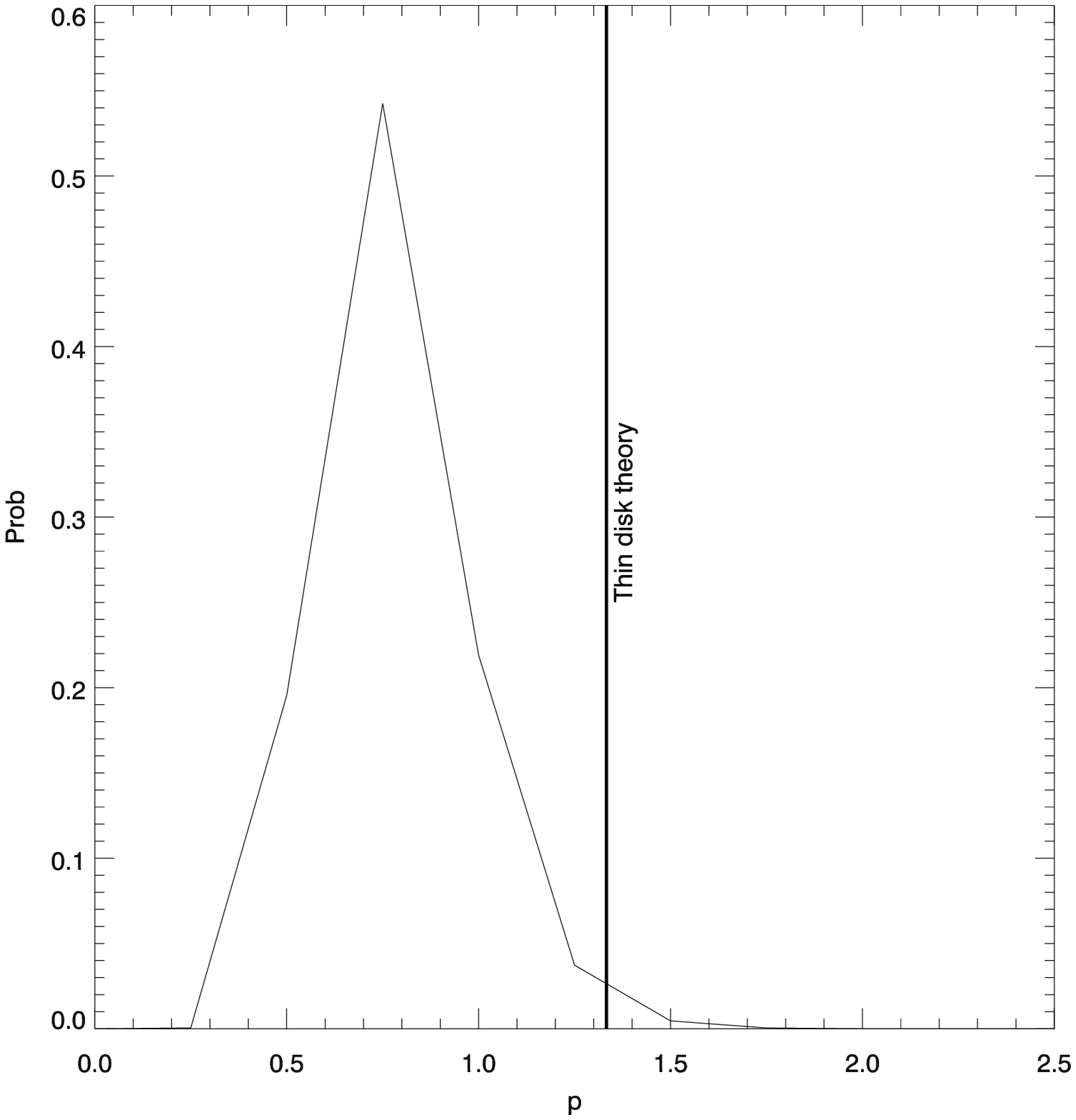}
\caption{Marginalized probability distribution for the dependence of disk size on wavelength $p$ ($r\propto \lambda^p$). 
The vertical line indicates the prediction of the thin disk standard model $p=4/3$. \label{fig4}}
\end{figure}
Figure \ref{fig4} shows the Bayesian estimate of $p$ found for our sample after
marginalizing over $\ln(r_0)$. We find $p=0.8\pm0.2$, with the value of the 
standard thin disk model ($p=4/3$) being
excluded at the
98\% confidence level. The significance drops to 90\% when 
objects with only narrow band data are excluded. 

\section{Conclusions}

We have used microlensing measurements for 10 image pairs from 8 lensed quasars
to study the structure of their accretion disks. By using
spectroscopic and narrow band photometric data 
we can eliminate the effects of differential extinction, which
can be
mistaken with chromatic microlensing and minimize any contamination from the
weakly microlensed emission lines that can reduce the 
microlensing amplitude and potentially alter
its dependence on wavelength.
We find a best estimate of the wavelength dependence of the accretion disk size of
$p=0.75^{+0.2}_{-0.2}$ for a maximum likelihood analysis, and $p=0.8\pm 0.2$ for a Bayesian analysis ($r_s\propto \lambda^p$).
This estimate is significantly smaller than the $p=4/3$ value predicted by the 
Shakura \& Sunyaev (1973) thin
disk model with a temperature profile of $T\propto r^{-3/4} \propto r^{-1/p}$.
Thus, our result seems to indicate a steeper temperature profile for the
accretion disk.

Agol \& Krolik (2000) proposed a model in which the magnetic connection between the black hole and
the disk is able to spin the disk up leading to a steeper temperature profile with $p=8/7$, which is more consistent with
our results. Another alternative is that there is some other structure on top of the accretion
disk.
Along this line, Abolmasov \& Shakura (2012), proposed that super Eddington accretion in the disks of quasars would generate an 
optically thick spherical envelope that could explain the lower values of the exponent $p$ 
found by Blackburne et al. (2011a), although they need an ad-hoc mechanism to
explain the small sizes of the X-ray-emitting regions of microlensed quasars.
Note, however, that Dai et al. (2010) find that the scattered light fraction
has to be very high ($\gtrsim 50\%$) before microlensing size estimates are
significantly shifted by scattering.
Abolmasov \& Shakura (2012) proposed that this mechanism would only work for the quasars with lower
black hole masses, although we do not find any trend of the parameter $p$ with the mass of the black hole.
In this line, Fukue \& Iino (2010) found that a fully thermalized relativistic spherical
wind around the black hole would show a profile of the type $T(R)\propto R^{-1}$ which is very similar to the produced by our analysis.
Note that such steeper temperature profiles make it more difficult to
reconcile the larger disk sizes found by microlensing. Solutions
to this problem would favor shallower profiles closer to a strongly irradiated
disk with $p=2$ (see the discussion in Morgan et al., 2010).
Our estimate for the size of the accretion disk of $r_s=4.5
$ lt-days (at 1026 \AA\  in the rest frame) is in agreement 
with other recent determinations and considerably larger than the expected value from a thin disk radiating as a black body. 

The large slope of the temperature profile of quasar accretion disks
estimated in this work makes the already serious problem posed
by the large sizes determined by microlensing for these objects even
worse.
As scattering does not seem to be the answer to this discrepancy,
and other suitable theoretical models are lacking, evidence
seems to indicate that
some key ingredient in the emission mechanism of accretion disks
in these objects may be missing in our present understanding. 
On the observational side, 
optical/infrared spectroscopy (or even better 
spectroscopic monitoring) for a larger sample of lensed quasars 
would be desirable to 
unambiguously confirm the large sizes and
slopes of the temperature profile of quasar accretion disks.
In the meantime, on the theoretical front, a thorough revision 
of the standard model
of accretion disks in quasars may also be necessary.

\acknowledgments
  
We gratefully acknowledge the anonymous referee for
useful suggestions that improved the presentation of
this work.
This research was supported by the Spanish Mi\-nis\-te\-rio de 
Educaci\'{o}n y Ciencia with the grants AYA2011-24728, AYA2007-67342-C03-01/03, AYA2010-21741-C03/02. J.J.V. is also supported by the Junta de Andaluc\'{\i}a through the FQM-108 project. J.A.M. is also supported by the Generalitat Valenciana with the
grant PROMETEO/2009/64. C.S.K. is supported by NSF grant AST-1009756.
V.M. gratefully acknowledge support from FONDECYT 1120741.
\,

\end{document}